\documentstyle[aps,epsfig]{revtex}

\begin{document}

\draft
\title{Vortex structure in mesoscopic superconductors}
\author{F. M. Peeters\cite{peeters} and B. J. Baelus}
\address{Departement Natuurkunde, Universiteit Antwerpen (UIA), Universiteitsplein 1,%
\\
B-2610 Antwerpen, Belgium}
\date{\today}
\maketitle

\begin{abstract}
The nonlinear Ginzburg-Landau equations are solved numerically in
order to investigate the vortex structure in thin superconducting
disks of arbitrary shape. Depending on the size of the system and
the strength of the applied magnetic field giant vortex,
multi-vortex and a combination of both of them are found. The
saddle points in the energy landscape are identified from which we
obtain the energy barriers for flux penetration and expulsion. \\
\textit{keywords:} vortex matter, mesoscopic, flux penetration,
flux quantization
\end{abstract}

\section{Introduction}

Recent advances in nanoscience have demonstrated that
fundamentally new physical phenomena are found when the size of
the sample shrinks and becomes comparable to the length scale of
the investigated phenomenon.

Superconductivity is a macroscopic quantum phenomenon and
therefore it is interesting to see how this quantum state is
influenced if the sample is reduced in size. For rings and hollow
cylinders this leads to the well-known Little-Parks effect
\cite{little} which results in a periodic variation of the
critical temperature as function of the applied magnetic field,
the period being determined by the magnetic flux value within the
tube.

Bulk superconductors are divided up into type-I ($\kappa <
1/\sqrt{2}$) and type-II ($\kappa >  1/\sqrt{2}$) superconductors,
the distinction between them is completely determined by the
Ginzburg-Landau parameter $\kappa=\lambda/\xi$, where $\lambda$ is
the magnetic field penetration depth and $\xi$ the coherence
length, which is a material parameter. The difference is clearly
seen in the magnetic response of the system, where type-I has a
complete Meissner effect while type-II superconductors can have a
state with partial expulsion of the field in which flux lines
penetrate the superconductor. The latter is a superconducting
state with disconnected circular areas of normal state. For
mesoscopic superconductors we find that type-I superconductors can
behave like type-II or even show some mixed behaviour depending on
the size of the system and therefore $\kappa$ is no longer the
only determining parameter characterizing the vortex state of the
system.

The vortex structure in superconductors will be described in the
framework of the Ginzburg-Landau (GL) theory, which consists of
two coupled nonlinear differential equations:

\begin{equation}
\left( -i\overrightarrow{\nabla }_{2D}-\overrightarrow{A}\right)
^{2}\Psi =\Psi \left( 1-\left| \Psi \right| ^{2}\right) ,
\end{equation}
\begin{equation}
-\Delta _{3D}\overrightarrow{A}=\frac{d}{\kappa ^{2}}\delta \left(
z\right) \overrightarrow{j}_{2D},
\end{equation}
where
\begin{equation}
\overrightarrow{j}_{2D}=\frac{1}{2i}\left( \Psi ^{\ast }\overrightarrow{%
\nabla }_{2D}\Psi -\Psi \overrightarrow{\nabla }_{2D}\Psi ^{\ast
}\right) -\left| \Psi \right| ^{2}\overrightarrow{A},
\end{equation}
is the density of superconducting current and $d$ the thickness of
the sample. The superconducting
wavefunction
satisfies the boundary conditions $\left. \left( -i\overrightarrow{\nabla }%
_{2D}-\overrightarrow{A}\right) \Psi \right| _{n}=0$ normal to the
sample
surface and $\overrightarrow{A}=\frac{1}{2}H_{0}\rho \overrightarrow{e}%
_{\phi }$ far away from the superconductor. The distances are
measured in units of the coherence length
$\xi=\hbar/\sqrt{-2m^{\ast}\alpha}$, the order parameter in $\Psi
_{0}=\sqrt{-\alpha/\beta}$ and the vector potential in
$c\hbar/2e\xi$. $\kappa=\lambda/\xi$ is the Ginzburg-Landau
parameter, and $\lambda =c\sqrt{m/\pi}/4e\Psi_{0}$ is the
penetration length. The magnetic field is
measured in $H_{c2}=c\hbar/2e\xi^{2}=\kappa\sqrt{2}H_{c}$, where $H_{c}%
=\sqrt{4\pi\alpha^{2}/\beta}$ is the critical field. Temperature
is included in $\xi$, $\lambda$, $H_{c2}$, through their
temperature dependencies: $\xi(T)={\xi(0)}/{\sqrt{|
1-T/T_{c0}|}}$, $\lambda(T)={\lambda(0)}/{\sqrt{| 1-T/T_{c0}| }}$,
$H_{c2}(T)=H_{c2}(0)| 1-{T}/{T_{c0}}|$, where $T_{c0}$ is the
critical temperature at zero magnetic field.

Different approaches have been used to solve the GL-equations. In
the lowest Landau level approximation \cite{palacios} a linear
combination of solutions of the linearized first GL-equation is
used and the internal (taken homogeneous) magnetic field and the
expansion coefficients are determined by minimizing the free
energy. An extension of this approach beyond the lowest Landau
level was given in Ref. \cite{schweigert1}. It was shown recently
\cite{PRB64} that in order to account for demagnetization effects
in the lowest Landau level approximation it is necessary to
introduce an effective $\kappa$ which depends on the size of the
superconductor. Trial functions \cite{Pogosov} with several
variational parameters were used to study the vortex configuration
in mesoscopic cylinders with suppressed surface superconductivity.
Here, we will solve the two GL equations numerically using a
finite difference technique. Details of this approach can be found
in Ref. \cite{PRB57}.

\section{Giant vortex to multi-vortex transition}
First we consider a circular disk with thickness $d$. Because of
the circular symmetry we assume that the order parameter is
cylindrical symmetric: $\Psi(x,y) = F(\rho)e^{-L\phi}$ where $L$
is the vorticity. In doing so we restrict our set of solutions to
a subset in which the modulus of the local order parameter is
axially symmetric.

Because of the non-linear term in the GL-equation the
superconducting state is in general non-axial symmetric even if
the sample is circular symmetric. A well-known example is the
Abrikosov lattice which has triangular symmetry. We found that for
our circular disks such non-axial symmetric states are found when
the radius of the disk is sufficiently large and when the magnetic
field is not too large. For small disks the boundary condition
dominates which imposes its symmetry on the order parameter. For
large magnetic fields the inner part of the disk becomes normal
and the superconducting state survives only near the sample
surface where the shape of it will again determine the symmetry of
the order parameter. The free energy of the superconducting disk
is shown in Fig. 1(a) as function of the magnetic field for
different values of the vorticity. The full curves are for the
giant vortex state and the dashed curves for the multi-vortex
states. The transition point between them is given by the open
dot. Figs. 1(c,d,e) shows a contour plot of the modulus of the
superconducting density for $L=3$ at $H_{0}/H_{c2}=0.62$, $0.72$
and $0.82$, respectively. Notice that with increasing magnetic
field the size of the vortices grows and they move closer to each
other. At the transition field the single vortices coalesce into
one giant vortex. This is a continuous transition and therefore of
second order \cite{prl98}.

The magnetization is shown in Fig. 1(b) where the vertical lines
indicate the ground state transitions. Notice that for small
magnetic fields we have a linear $M-H$ relation which is typical
for an ideal diamagnet. With increasing magnetic field there is a
{\it continuous} penetration of the magnetic field at the edge of
the sample which leads to a smaller than linear increase of $M$
with $H$. This effect is enhanced by demagnetization effects which
leads to an enhanced magnetic field at the edge of the sample (see
inset of Fig.~1(a)). When we further increase the magnetic field
the energy of the superconductor increases up to the point where
it is energetically more favorable to transit to the $L=1$ state.
As a consequence $M(H)$ exhibits a zig-zag behavior which was
measured recently by A. Geim {\it et al} \cite{geim97} and
explained in Ref. \cite{deo97}. This behavior is due to the fact
that with increasing magnetic field more field can penetrate into
the sample in a {\it discontinuous} way in which the vorticity of
the order parameter increases with one unit. Approximately
(because of boundary effects), one unit of flux enters the
superconductor. Therefore, by counting the number of jumps in the
$M(H)$ curve it is possible to count the vorticity of the sample
and therefore the number of vortices inside the disk.

Notice that for a given magnetic field different states are
possible. Experimentally, it has been possible to drive the system
into the different metastable regions and to map out the complete
magnetization-magnetic field curves as shown in Fig. 1(b) by
slowly ramping the field up and down. These metastable states are
responsible for hysteretic behavior which was observed in the
magnetic field response of superconducting disks. A detailed
comparison between experimental results and our theoretical
calculations can be found in Ref. \cite{deo98}. It was shown
recently \cite{geim00} that in certain samples the superconducting
state can be forced very far into the metastable region such that
fractional flux can penetrate and in some cases even negative flux
penetration is observed for the transition $L \rightarrow L+1$.

\section{Geometry dependence}

One may wonder how important the exact shape of the sample is for
the symmetry of the order parameter. To investigate this, we took
as an example a thin flat triangular shaped superconducting
sample. For the free energy we found a similar behavior as
depicted in Fig. 1(a) but with the following major differences: 1)
the critical magnetic field $H_{c3}$ at which the normal state is
reached is substantially larger, i.e. for a triangle with the same
surface area as the disk of Fig. 1(a) we obtained
$H_{c3}/H_{c2}=2.5$. This is a consequence of the enhanced surface
conductivity in wedge shaped samples \cite{PRB60,Misko}; 2) The
multi-vortex state (see Fig. 2(a)) is substantially stabilized,
i.e. the magnetic field range over which it is stable is strongly
enhanced. Furthermore, for certain values of the vorticity only
the multi-vortex state is stable, and the giant vortex state does
not occur. The giant vortex state is now no longer circular
symmetric, but it is characterized by the fact that the order
parameter has only one zero. 3) A new vortex structure appears
which is a combination of a giant vortex in the center of the
triangle with single vortices around it. This state occurs e.g.
for $L=5$; 4) From the contour plot of the superconducting current
flow (see Fig. 2(b)) we see that the screening currents along the
edge flow clockwise while the currents around the vortices flow
counterclock wise. Notice that near the edges their are clockwise
spiralling local vortex-like currents which are not connected to
the position of a vortex (see the contour plot of Fig. 2(c) for
the phase of the order parameter) but are rather back-flow
currents which are well-known in hydrodynamics.

Recently, Chibotaru and co-workers \cite{Chibotaru} studied the
linearized GL-equation for triangles and squares and found that
the symmetry of the boundary imposes a similar symmetry on the
superconducting state. For example, in case of the triangle and
for $L=2$ this occurs by having three vortices sitting in a
triangular arrangement and one anti-vortex in the center of this
triangle. From the outset we should point out that the linearized
GL-equation is only strictly valid at the superconducting/normal
boundary where the modulus of the order parameter approaches zero.
It was found recently \cite{bonca} that once one moves away from
this phase line the anti-vortices are quickly annihilated. Then a
state is found in which a large central area of the triangle is
normal and only superconductivity survives in the corners of the
triangle \cite{baelus01}. The anti-vortices and the vortices are
found in the central area of the triangle where $|\Psi|^2$ is
extremely small. Therefore, the experimental relevance of this new
state is questionable.

\section{Different arrangements of multi-vortices}

When the superconducting state is in the multi-vortex state, it is
possible that different vortex configurations are possible. This
is illustrated in Fig. 3 where for a disk of radius $R/\xi=6$ the
energy of the $L=6$ and $L=7$ vortex states are plotted as
function of the magnetic field. In this case a multi-vortex
configuration is possible with (dashed curves) or without (solid
curves) a vortex in the center of the disk. The  corresponding
contourplots of the superconducting density in the disk is shown
in the insets of Fig. 3. The states are indicated by $(n;L)$ where
$n$ refers to the number of vortices in the center and $L$ is the
vorticity of the superconducting state. Notice that for $L=6$ the
(1;6) state has a larger energy than (0;6) when $H/H_{c2} > 0.45$
which then becomes the ground state. This change of configuration
at $H/H_{c2} \sim .45$ is a first order transition because it
involves a change of symmetry of the superconducting state. At
$H/H_{c2} \sim 0.6$ the ground state transits to a higher
vorticity state with configuration (1;7). This transition is also
a first order transition in which the magnetization is
discontinuous. Such transitions have been found experimentally
\cite{geimprl}.

Notice that the appearance of the different multi-vortex
configurations is very similar to the classical system of
interacting (repulsion) particles which are confined into a
potential. It was found that the particles are situated in
ring-like configurations \cite{bedanov}, very similar to the
vortex configurations found in Fig. 3. Also in this case different
meta-stable states consisting of different particle configurations
are found.

\section{Flux penetration and expulsion}

In the above discussion we considered only the local minima in the
energy functional which determine the ground and metastable
states. The transition between states with different vorticity
does not always follow the ground state. The reason is that their
are barriers for flux penetration and explusion. The most
well-known is the Bean-Livingston model \cite{bean} for the {\it
surface barrier} which is a result of the competition between the
vortex attraction to the sample walls by its mirror image and its
repulsion by screening currents. For nonelliptical samples a {\it
geometrical barrier} appears because of Meissner currents flowing
on the top and bottom surface \cite{brandt}. Additionally, {\it
vortex pinning} by defects can play an important role in the delay
of vortex expulsion or promotion of vortex penetration. These
models describe the vortex formation far from the sample boundary.

As an example, we will consider here mesoscopic disks in which
boundary effects are predominant and consequently previous
approaches are not applicable. We consider a defect-free thin
superconducting disk with a perfect circular boundary such that
$Rd/\lambda^2 \ll 1$. When calculating the free energy we not only
search for the local minima but also for the saddle points which
are the lowest energy barriers between two such minima. The
details of the calculation can be found in Refs.
\cite{schweigert1,baelusprb}.

The results are shown in Fig. 4 where the energy of the saddle
point is given by the dotted curve. In contrast to known surface
and geometrical barrier models, we find that in a wide range of
magnetic fields below the penetration field, the saddle point
state for flux penetration into a disk does not correspond to a
vortex located near the sample boundary, but to a region of
suppressed superconductivity (see Figs. 4(a-d)) at the disk edge
with no winding of the current (Fig. 4(c)), and which is a {\it
nucleus} for the following vortex creation. The height of this
{\it nucleation barrier} is shown in the inset of Fig. 4 and
determines the time of flux penetration.

\section{Conclusions}

The vortex state of a mesoscopic superconductor is strongly
determined by its size and to a lesser extend by the material
parameters the superconductor is made of. For example, by
increasing the radius (starting from $R \ll \xi$) of a
superconducting disk \cite{geim97,deo97} it is possible to obtain
a magnetic response which, as function of the magnetic field, is
continuous, type-I, a type-I with multiple steps and type-II. The
exact geometry of the superconductor has also a strong influence
on its vortex state. This vortex state can be brought into a
metastable region owing to the presence of barriers for flux
motion which leads to unexpected effects like fractional flux and
even negative flux entry. In mesoscopic superconductors the flux
in general is not quantized \cite{geim00}. This is even more so in
small superconducting rings \cite{baelus00}. The lowest energy
barrier between two flux states was identified and corresponds to
a saddle point of the energy functional.

\section*{Acknowledgments}

This work was supported by the Flemish Science Foundation
(FWO-Vl), the ''Onderzoeksraad van de Universiteit Antwerpen''
(GOA), the ''Interuniversity Poles of Attraction Program - Belgian
State, Prime Minister's Office - Federal Office for Scientific,
Technical and Cultural Affairs'', and the European ESF programme
on 'Vortex Matter'. Discussions with S. Yampolskii, V. Schweigert,
V. Moshchalkov, A. Geim and L. Chibotaru are gratefully
acknowledged.

\begin{figure}[tb]
\caption{(a) The free energy and (b) the magnetization of the
multi-vortex states (dashed curves) and the giant vortex states
(solid curves) as a function of the applied magnetic field for a
disk with radius $R=4.0\xi$ and thickness $d=0.1\xi$
($\kappa=0.28$). The inset in (a) shows the radial magnetic field
distribution through the central plane of the disk for the $L=3$
state at $H_{0}/H_{c2}=0.82$ (giant vortex state). The open
circles indicate the transition from multi-vortex to giant-vortex
states. (c-e) Contour plots of the Cooper-pair density for the
$L=3$ state at $H_{0}/H_{c2}=0.62$, $0.72$ (multi-vortex states)
and $0.82$ (giant vortex state). High (low) Cooper-pair density is
given by dark (light) regions.}
\end{figure}

\begin{figure} [tb]
\caption{(a) The Cooper-pair density, (b) the supercurrent and (c)
the phase of the order parameter for a triangle with width
$W=10.774\xi$ and thickness $d=0.1\xi$ ($\kappa=0.28$). The
applied field is $H_{0}=0.495H_{c2}$ and $L=2$.}
\end{figure}

\begin{figure} [tb]
\caption{The free energy $F$ as a function of the applied magnetic field $%
H_{0}$ of the $(0;6)$ and $(0;7)$ state (solid curves), and the
$(1;6)$ and $(1;7)$ state (dashed curves) for a superconducting
disk with radius $R=6.0\protect\xi$. The insets show the
Cooper-pair density of the four different states at
$H_{0}/H_{c2}=0.6$.}
\end{figure}

\begin{figure}[tb]
\caption{The free energy as a function of the applied magnetic
field $H_0$ for a circular disk of radius $R/\xi=4$. The solid
curves correspond to the giant vortex state, the dashed curves are
the multi-vortex states and the dotted curves are the energy of
the saddle point. The inset shows the lowest energy barrier for
the transition $L \rightarrow L+1$. (a-d) show contour plots of
the superconducting density for the saddle point corresponding to
the $L=4 \rightarrow 5$ transition at the magnetic fields
$H_0/H_{c2}$=0.81, 0.885, 0.96 (the barrier maximum), 1.035,
respectively.}
\end{figure}


\begin{references}
\bibitem[%
\circ%
%
]{peeters}  Electronic mail: peeters@uia.ua.ac.be
\bibitem{little}
W.A. Little and R.D. Parks, Phys. Rev. Let. {\bf 9}, 9 (1962);
{\it ibid.} Phys. Rev. {\bf 133}, A97 (1964).
\bibitem{palacios}
J.J. Palacios, Phys. Rev. B {\bf 58}, R5948 (1998); S.V. Yampolskii and
F.M. Peeters, Phys. Rev. B {\bf 62}, 9663 (2000).
\bibitem{schweigert1}
V.A. Schweigert and F.M. Peeters, Phys. Rev. Lett. {\bf 83}, 2409 (1999).
\bibitem{PRB64}
J.J. Palacios, F.M. Peeters, and B.J. Baelus, Phys. Rev. B {\bf
64}, 134514 (2001).
\bibitem{Pogosov}
W.V. Pogosov, cond-mat/0108492.
\bibitem{PRB57}
V.A. Schweigert and F.M. Peeters, Phys. Rev. B {\bf 57}, 13817
(1998).
\bibitem{prl98}
V.A. Schweigert, F.M. Peeters, and P.S. Deo, Phys. Rev. Lett. {\bf
81}, 2783 (1998).
\bibitem{geim97}
A.K. Geim, I.V. Grigorieva, S.V. Dubonos, J.G.S. Lok, J.C. Maan,
A.E. Filippos, and F.M. Peeters, Nature (London) {\bf 390}, 256
(1997).
\bibitem{deo97}
P.S. Deo, V.A. Schweigert, F.M. Peeters, and A.K. Geim, Phys. Rev.
Lett. {\bf 79}, 4653 (1997); P.S. Deo, F.M. Peeters, and A.K.
Geim, Superlattices Microstr. {\bf 25}, 1195 (1999).
\bibitem{deo98}
P.S. Deo, V.A. Schweigert, and F.M. Peeters, Phys. Rev. B {\bf
59}, 6039 (1999).
\bibitem{geim00}
A.K. Geim, S.V. Dubonos, I.V. Grigorieva, K.S. Novoselov, F.M.
Peeters, and V.A. Schweigert, Nature (London) {\bf 407}, 55
(2000).
\bibitem{PRB60}
V.A. Schweigert and F.M. Peeters, Phys. Rev. B {\bf 60}, 3084
(1999).
\bibitem{Misko}
V.R. Misko, V.M. Fomin, and J.T. Devreese, Phys. Rev. B {\bf 64},
014517 (2001).
\bibitem{Chibotaru}  L.F. Chibotaru, A. Ceulemans, V. Bruyndoncx and V.V.
Moshchalkov, Nature (London) {\bf 408}, 833 (2000); {\it ibid.},
Phys. Rev. Lett. {\bf 86}, 1323 (2001).
\bibitem{bonca}
J. Bon\v ca and V.V. Kabanov, Phys. Rev. B {\bf 65}, 012509
(2001).
\bibitem{baelus01}
B.J. Baelus and F.M. Peeters, cond-mat/0106601.
\bibitem{geimprl}
A.K. Geim, S.V. Dubonos, J.J. Palacios, I.V. Grigorieva, M.
Henini, and J.J. Schermer, Phys. Rev. Lett. {\bf 85}, 1528 (2000);
V.A. Schweigert and F.M. Peeters, Physica C {\bf 332}, 266 (2000).
\bibitem{bedanov}
V.M. Bedanov and F.M. Peeters, Phys. Rev. B {\bf 49}, 2667 (1994);
B. Partoens and F.M. Peeters, J. Phys.: Condens. Matter {\bf 9}, 5383 (1997);
M. Kong, B. Partoens, and F.M. Peeters, cond-mat/0106395, to appear in Phys. Rev. E {\bf 65} (2002).
\bibitem{bean}
C.P. Bean and J.B. Livingston, Phys. Rev. Lett. {\bf 12}, 14 (1964).
\bibitem{brandt}
Th. Schuster, M.V. Indenbom, H. Kuhn, E.H. Brandt, and M. Konczykowski,
Phys. Rev. Lett. {\bf 73}, 1424 (1994).
\bibitem{baelusprb}
B.J. Baelus, F.M. Peeters, and V.A. Schweigert, Phys. Rev. B {\bf 63}, 144517 (2001).
\bibitem{baelus00}
B.J. Baelus, F.M. Peeters, and V.A. Schweigert, Phys. Rev. B {\bf 61}, 9734 (2000).
\end{references}
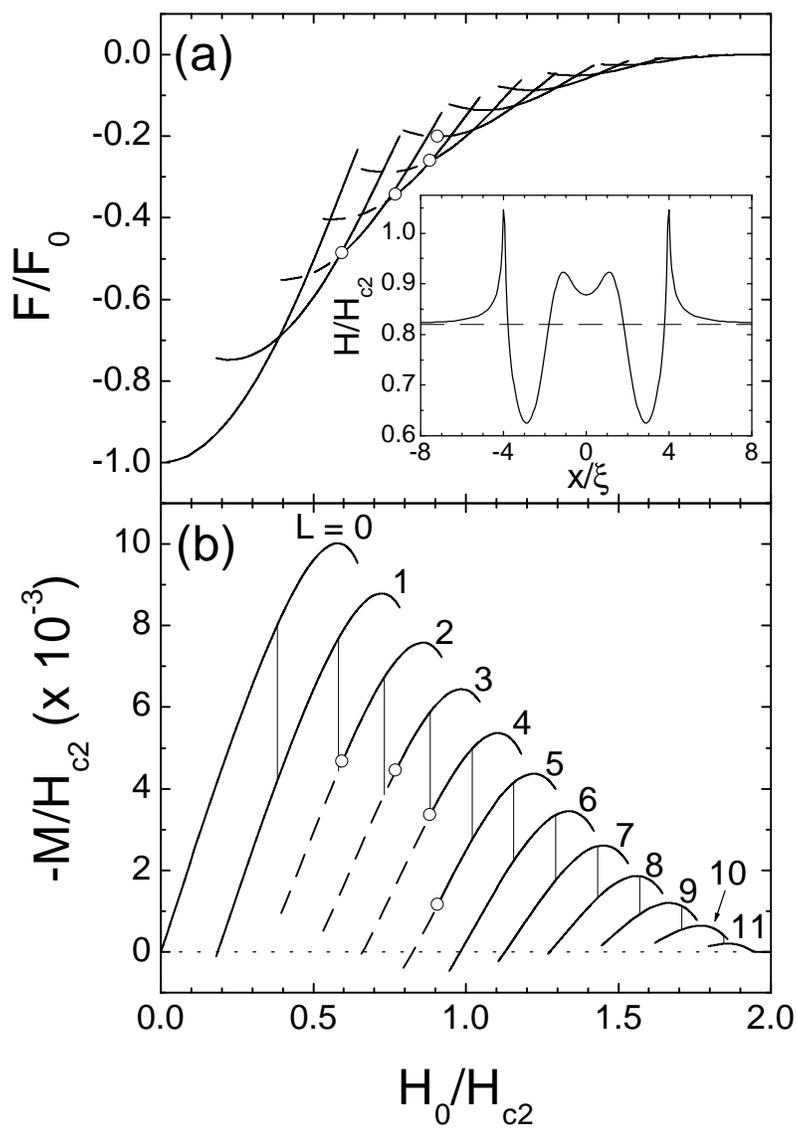
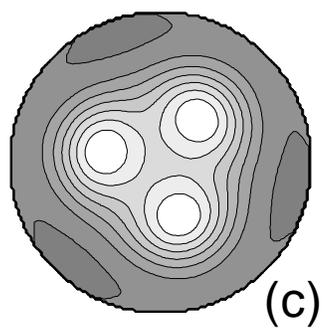
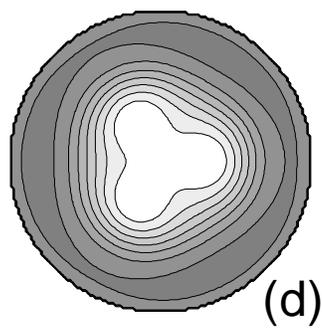
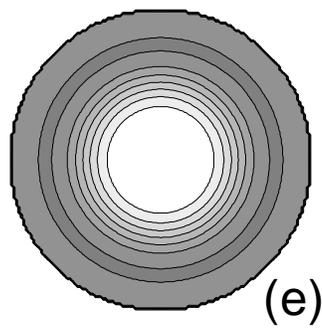
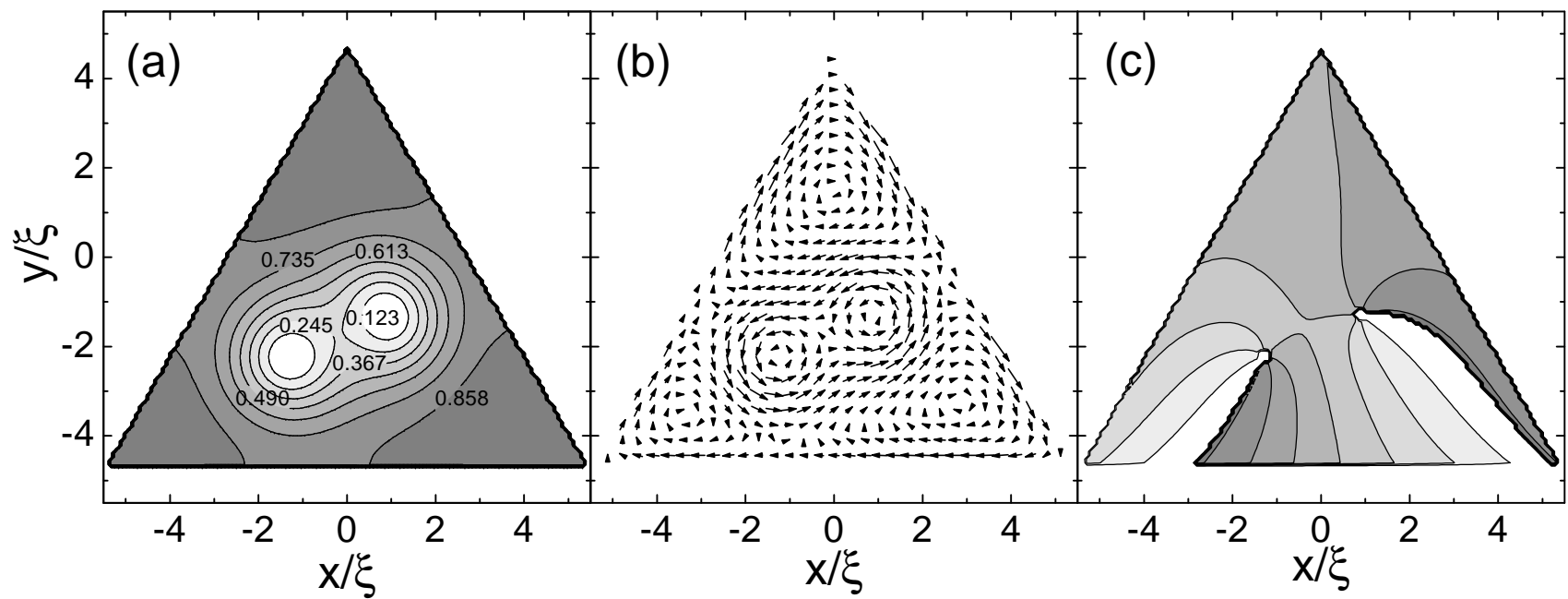
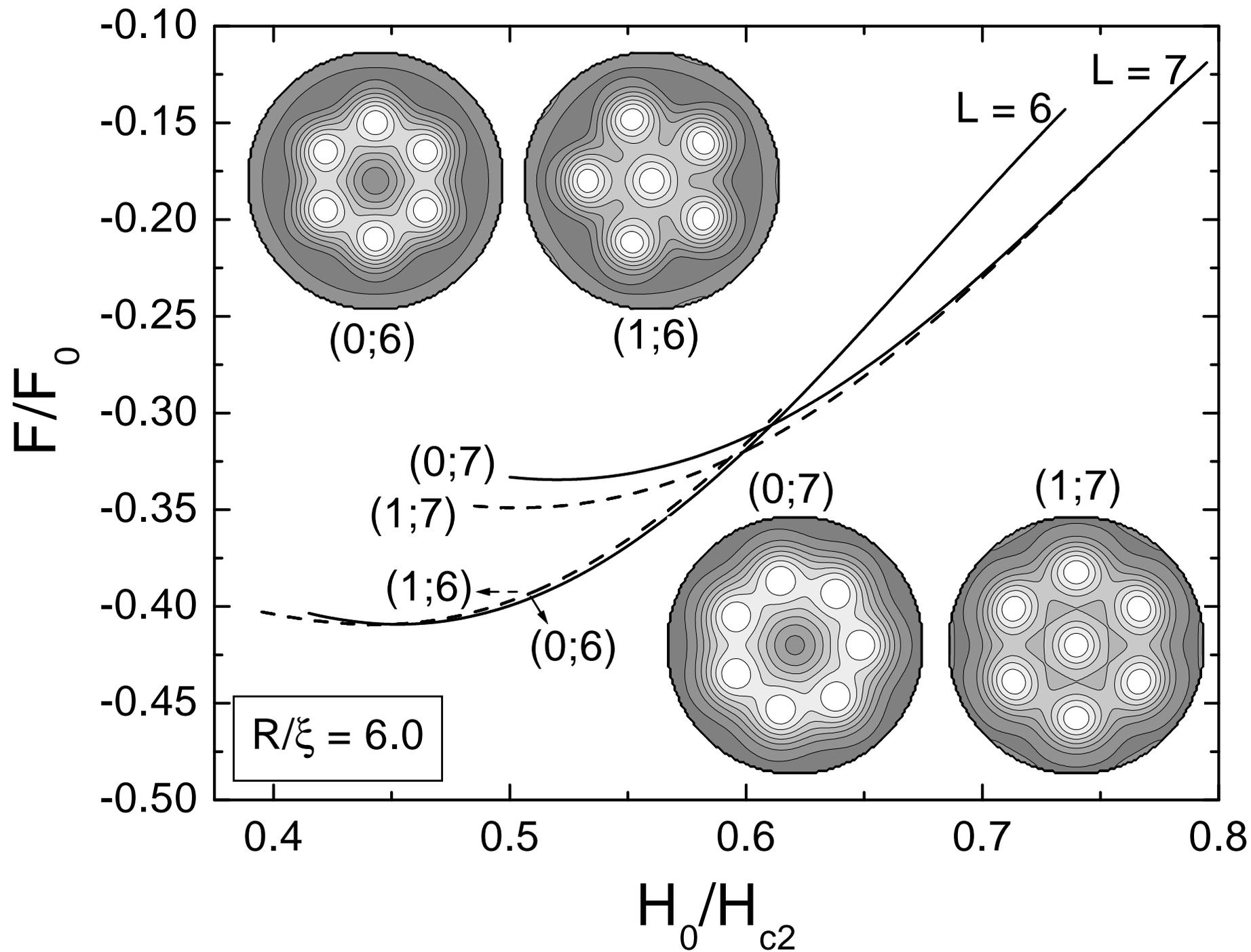
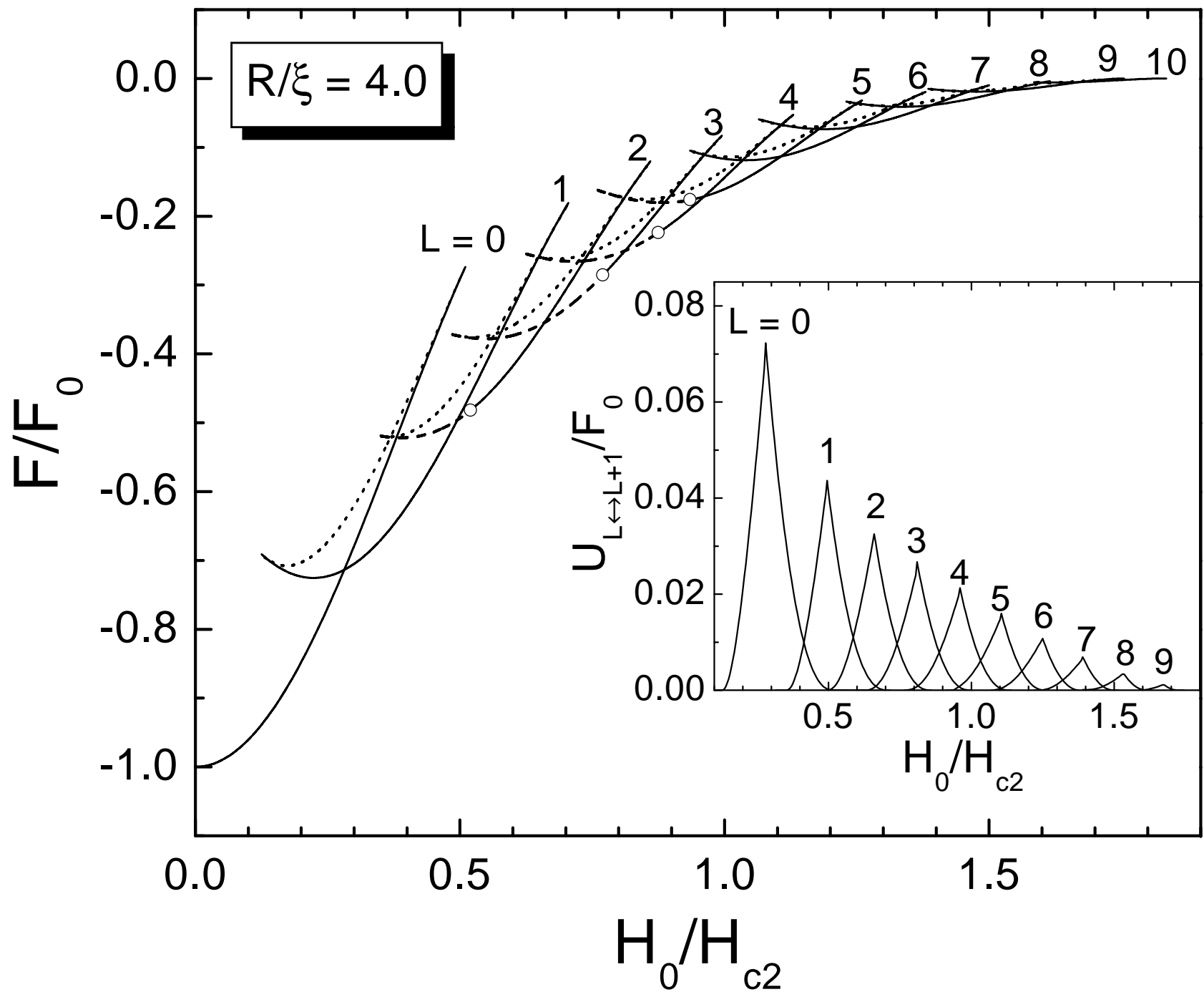
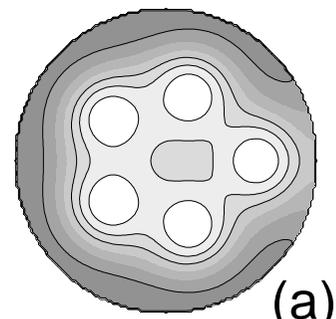
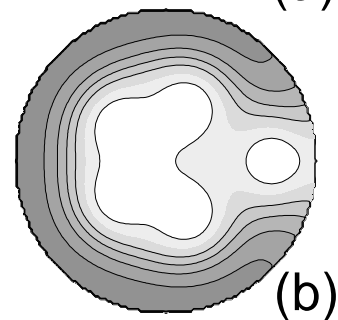
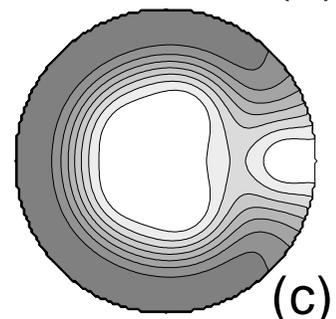
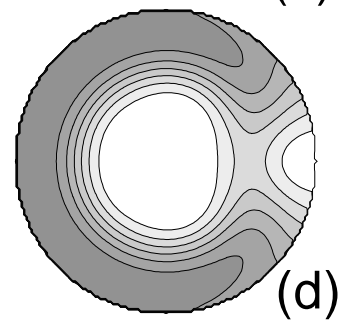
\end{document}